\documentclass[sigconf]{acmart}

\usepackage{threeparttable}

\AtBeginDocument{%
  }

\setcopyright{rightsretained}
\copyrightyear{2024}
\acmYear{2024}
\acmConference[Genaiecom '24]{Proceedings of the first workshop on Generative AI for E-Commerce}{October 25, 2024}{Boise, ID}
\acmBooktitle{Proceedings of the first workshop on Generative AI for E-Commerce 2024, October 25, 2024}
\acmYear{2024}
\acmMonth{10}
\acmDOI{}
\acmISBN{}
\acmPrice{15.00}

\begin{document}

\title{Towards More Relevant Product Search Ranking Via \\ Large Language Models: An Empirical Study \\ \ }

\author{Qi Liu}
\email{qi.liu@walmart.com}
\affiliation{%
  \institution{Walmart Global Tech}
  \city{Hoboken}
  \state{NJ}
  \country{USA}
}

\author{Atul Singh}
\email{atul.singh@walmart.com}
\affiliation{%
  \institution{Walmart Global Tech}
  \city{Hoboken}
  \state{NJ}
  \country{USA}
}

\author{Jingbo Liu}
\email{jingbo.liu@walmart.com}
\affiliation{%
  \institution{Walmart Global Tech}
  \city{Hoboken}
  \state{NJ}
  \country{USA}
}

\author{Cun Mu}
\email{cun.mu@walmart.com}
\affiliation{%
  \institution{Walmart Global Tech}
  \city{Hoboken}
  \state{NJ}
  \country{USA}
}

\author{Zheng Yan}
\email{zheng.yan0@walmart.com}
\affiliation{%
  \institution{Walmart Global Tech}
  \city{Hoboken}
  \state{NJ}
  \country{USA}
}


\addtolength{\topskip}{0.35cm}

\begin{abstract}
Training Learning-to-Rank models for e-commerce product search ranking can be challenging due to the lack of a gold standard of ranking relevance. In this paper, we decompose ranking relevance into content-based and engagement-based aspects, and we propose to leverage Large Language Models (LLMs) for both label and feature generation in model training, primarily aiming to improve the model's predictive capability for content-based relevance. Additionally, we introduce different sigmoid transformations on the LLM outputs to polarize relevance scores in labeling, enhancing the model's ability to balance content-based and engagement-based relevances and thus prioritize highly relevant items overall. Comprehensive online tests and offline evaluations are also conducted for the proposed design. Our work sheds light on advanced strategies for integrating LLMs into e-commerce product search ranking model training, offering a pathway to more effective and balanced models with improved ranking relevance.
\end{abstract}



%%
%% Keywords.


\maketitle

\section{Introduction}
\label{sec:intro}

As an essential daily activity for millions of users, online shopping has become a crucial element of people’s lives. Modern e-commerce product search platforms handle a vast multitude of customer search queries, striving to deliver the most relevant and deserving products in the search results. Product search ranking is one of the most crucial components in the search tech stack, where advanced machine learning models are employed to present online customers with highly relevant products for their search queries. With the rapid advancements in generative artificial intelligence (GenAI) methodologies, particularly large language models (LLMs), there is significant potential to leverage these technologies to improve the relevance of search ranking results. This, in turn, can enhance the system's ability to promote higher-quality products to top positions.

The application of pre-trained language models in search ranking modeling has been widely explored in previous works. For example, as one of the most popular choices, BERT \cite{bert} has been extensively used in encoding queries and documents to capture and evaluate semantic relatedness in many studies \cite{nogueira2019passage, nogueira2019multi, lin2020pre, li2021lightweight, jiang2021bert}, which achieve state-of-the-art performance on various benchmarks optimizing for ranking relevance. Similar research on search ranking was also performed leveraging variant models of BERT, e.g., DistilBERT \cite{sanh2019distilbert}, ALBERT \cite{lan2020albert}, RoBERTa \cite{liu2019roberta}. With the emergence of more complex LLMs such as LlaMa \cite{llama1, llama2}, GPT \cite{gpt3, gpt4}, Mistral \cite{mistral}, and Gemma \cite{gemma2024open}, which are trained on extensive datasets with a larger number of parameters, offering greater flexibility and generalization, and demonstrating improved performance across various semantic tasks \cite{kaplan2020scaling, liu2023large, holistic2023}, it is increasingly tempting to utilize them in search ranking tasks.

Most current methods for integrating LLMs into ranking model training focus on directly using query-product level predicted scores from LLMs as relevance labels for downstream tasks or student models \cite{Yao2022ReprBERTDB, Liu2023Milti, vo2024knowledge}. However, there is less discussion on how to further refine the LLM output for optimal use and how to better integrate LLMs into ranking model training to fully leverage their semantic capabilities. Our work addresses this gap by implementing several key strategies. On the one hand, we integrate LLMs of varying sizes into generating labels and features for ranking model training. While the labels can be generated using more complex language models, smaller models are used for features due to runtime latency concerns. This approach maximizes improvements in content-based relevance for ranking. On the other hand, we decompose the target for product ranking—relevance labels—into content and engagement components, leveraging LLMs specifically for content relevance. Additionally, we propose several sigmoid-based transformations on the LLM output to generate content relevance, with different transformation choices influencing the dynamics of the engagement vs. content trade-off.

The remainder of the paper is organized as follows: Section~\ref{sec:method} discusses our proposed methods for integrating LLMs into training search ranking models. Section~\ref{sec:experiments} details the empirical experiments and evaluations conducted to assess our proposals. Finally, Section~\ref{sec:conclusion} summarizes our findings and presents our conclusions.

\section{Methodology}
\label{sec:method}

% One challenge in ranking model training is that, unlike many other machine learning tasks where labels have a simple and unambiguous definition, there is no gold standard for generating ground-truth ranking labels. The root cause is the lack of a universal criterion for defining a perfect ranking model. However, this challenge also presents an opportunity to improve modeling, not only through the utilization of high-quality features but also by customizing the labels to guide the model in generating more relevant ranking results that suit specific goals for different use cases. We incorporate both techniques to enhance the relevance of ranking models.

Our search ranking models are trained using the Learning-to-Rank (LTR) framework \cite{liu2009, karmaker2017application, eletreby2022machine} optimized for search events \cite{Tong2022Patent} by utilizing data from a truncated historical period of online customer search traffic on Walmart.com, a major e-commerce platform. The improvement of ranking model training is achieved not only through the utilization of high-quality features but also by customizing the labels to guide the model in generating more relevant ranking results that suit specific goals for certain use cases. In this work, our primary focus is on optimizing the integration and utilization of LLMs into both the label and feature generation for model training.

\subsection{Label Formulation}
\label{subsec:label}

% We adopt the listwise approach \cite{cao2007} for our LTR modeling, where the loss function aims to optimize the Normalized Discounted Cumulative Gain (NDCG) metrics \cite{jarvelin2002cumulated}, with relevance scores playing a vital role in the calculation. In the context of e-commerce search, we decompose the ranking relevance into two major components: content-based relevance and engagement-based relevance. The former gauges how pertinent a product's attributes (brand, gender, color, product type, etc.) are to a search query, while the latter evaluates how frequently customers interact with a product based on their judgment under a search query. Accordingly, we propose the following label formulation: given a group $g$ of products $\{p_i\}_{i=1}^{|g|}$ to rank under search query $q$, we assign label

% % \begin{equation}
% %     y_{g, i} = \sigma\Big(C_{q, i}\Big) \cdot \eta^{\text{rank} \big(\sum_t E_{g_t, i}\big) \cdot \mathcal{O}\big(E_{g, i}\big)},
% % \end{equation}

% \begin{equation}\label{eqn:label}
%      y_{g, i} = \sigma\Big(C_{q, i}\Big) \cdot E\Big(g, q, i\Big),
% \end{equation}
% where $C_{q, i} \in [0,1]$ is an inferred content-based relevance score for query-product pair $(q, i)$, $\sigma(\cdot)$ is a transformation function for the content relevance, and $E(\cdot)$ yields the engagement-based relevance score, a value based on the engagement types (ordered $>$ added-to-cart $>$ clicked $>$ non-engaged) for product $p_i$ in the search session for group $g$ and historical search sessions associated with query $q$.

We adopt the listwise approach \cite{cao2007} for our LTR modeling, where the loss function aims to optimize the Normalized Discounted Cumulative Gain (NDCG) metrics \cite{jarvelin2002cumulated}, with relevance scores playing a vital role in the calculation. In the context of e-commerce search, we decompose the ranking relevance into two major components: content-based relevance and engagement-based relevance. The former gauges how pertinent a product's attributes--e.g., title, description, brand, gender, color, product type, etc.--are to a search query, while the latter evaluates how frequently customers interact with a product based on their judgment under a search query. Accordingly, we propose the following label formulation: given a group $\mathcal G$ of products $\{ p_1, p_2, \ldots, p_{|\mathcal G|}\}$, we assign label 
\begin{equation}\label{eqn:label}
     y^{\mathcal G}_{q, p} = \sigma\Big(C_{q, p}\Big) \cdot E^{\mathcal G} \Big(q, p\Big),
\end{equation}
where $C_{q, p} \in [0,1]$ is an inferred content-based relevance score for query-product pair $(q, p)$, $\sigma(\cdot)$ is a transformation function for the content relevance, and $E^{\mathcal G} (\cdot)$ yields the engagement-based relevance score, a value based on the engagement types (ordered $>$ added-to-cart $>$ clicked $>$ non-engaged) for product $p$ in the search event with query $q$ and product group $\mathcal G$. 

Compared to engagement-based relevance score $E$, which entirely relies on factual logged customer behavioral data and is therefore objective, content-based relevance $C$ is more subjective, as it depends heavily on semantic interpretation. Given the substantial volume of query-product pairs in e-commerce catalogs, using human-labeled datasets to train machine learning models to predict relevance scores becomes an appealing approach. In this specific instance, we propose to leverage LLM models, distinct from the ranking model, to generate content relevance scores due to their demonstrated exceptional performance on natural language tasks. Specifically, we fine-tune a Mistral 7B model \cite{mistral} using in-house human-evaluated data with the cross-entropy loss
\begin{equation}\label{eqn:llm_loss}
     \mathcal{L}_{q,p} = -r_{q,p}\log\Big(\hat{r}_{q,p}\Big) - \Big(1 - r_{q,p}\Big)\log\Big(1 - \hat{r}_{q,p}\Big),
\end{equation}
where $r_{q,p}$ is the human-evaluated content-based relevance label for query-product pair $(q, p)$ and $\hat{r}_{q,p}$ is the LLM predictions. The fine-tuned LLMs will then be able to infer content-based relevance scores $C_{q,p} \in [0,1]$ for any given $(q, p)$, where the closer the value is to 1, the more content-relevant the product is.

Though the multiplication of content and engagement can account for both aspects, it also exhibits a potential trade-off between the two, where the engagement performance lift in a ranking model could negatively impact its performance in content-based relevance and vice versa.

\subsection{Relevance Transformation}
\label{subsec:transformation}

Another perspective for comparing content-based vs. engagement-based relevances is that content reflects endogenous properties of a product, as it originates from inherent attributes of the product itself. In contrast, engagement represents exogenous characteristics, as it depends on how end users perceive and interact with the product. Hence, we let content relevance function as a guardrail, as an ideal ranking model should be able to distinctly differentiate products with significant content relevance gaps, ensuring that highly relevant ones are ranked in top positions. For products within the same content relevance interval, the model should primarily rely on user engagement to determine their ranking. Following this principle, we propose a sigmoid transformation on the content relevance score used in label~(\ref{eqn:label})
\begin{equation}\label{eqn:sigmoid}
     \sigma(C; \alpha, \beta) = \frac{1}{1 + \exp\big[ -\alpha(C - \beta) \big]},
\end{equation}
where $C$ is the LLM-inferred content relevance score, and $\alpha, \beta > 0$ are shape parameters determining the center and steepness of the sigmoid curve. The main rationale behind this transformation is that the sigmoid function can polarize those moderately low and high scores toward both extremes, effectively differentiating the content relevance gap between products. Additionally, within each interval at both ends, the content curve remains relatively flat, allowing engagement $E$ to play a more significant role in the label.

\begin{figure}
  \includegraphics[scale=0.18]{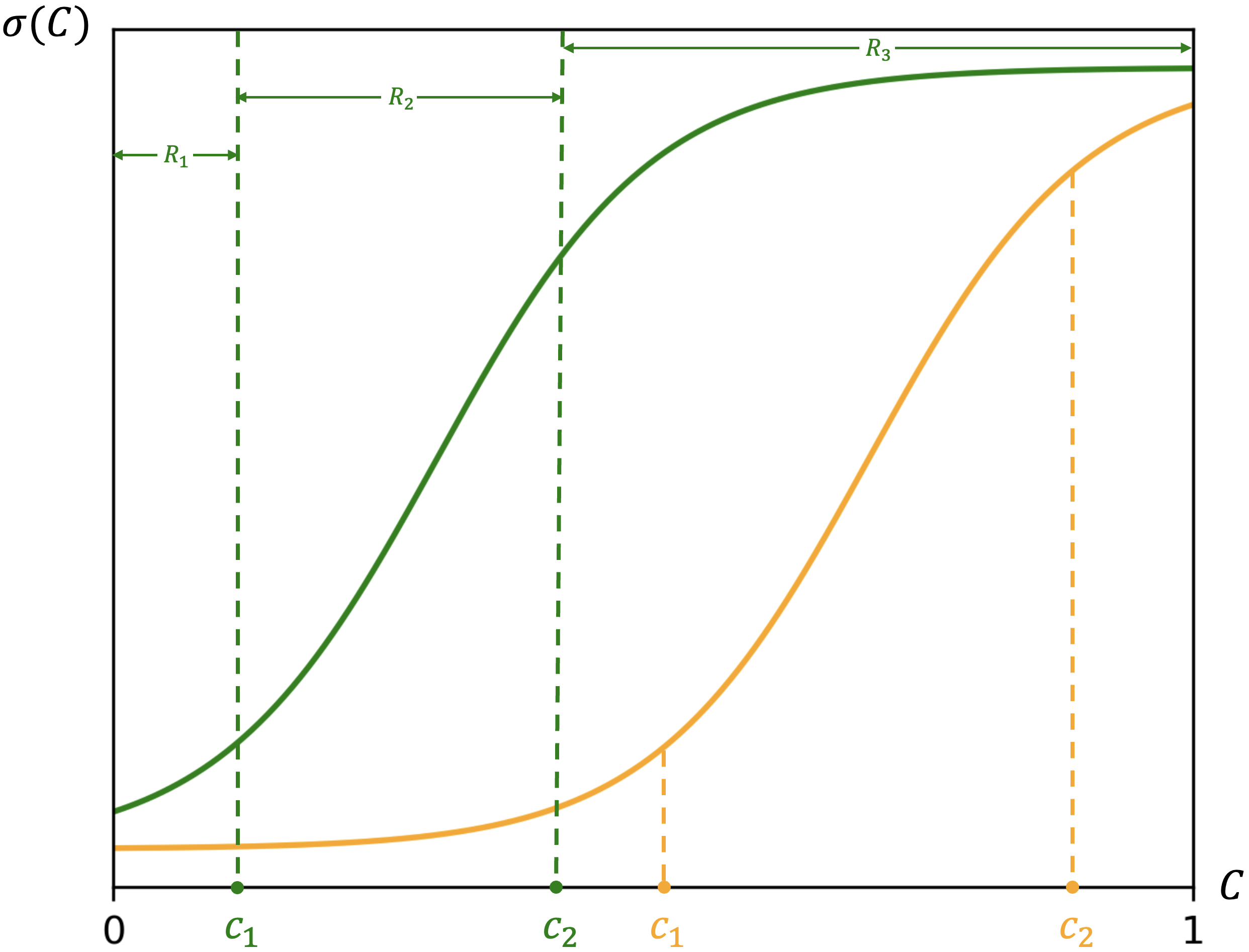}
  \caption{Curve of Sigmoid Transformation}
  \label{fig:curve}
\end{figure}

Shown in Figure~\ref{fig:curve} is the comparison between different sigmoid transformation curves. We segment query($q$)-product($p$) pairs (QPs) into three intervals $\{R_1, R_2, R_3\}$ according to their original content relevance scores $C_{q,p}$ predicted by LLMs. Suppose $0 < c_1 < c_2 < 1$,
\begin{itemize}
    \item $R_1 = \{(q,p): 0 \le C_{q,p} < c_1\ \text{s.t.}\ 0 < \nabla_{C_{q,p}} \sigma \le 1 \}$: QPs falling into this interval have low content relevance. They are flattened by the transformation and pushed to have even lower scores. The distinction among these QPs is more determined by their engagement $E^{\mathcal G} \big(q, p\big)$.
    \item $R_2 = \{(q,p): c_1 \le C_{q,p} < c_2\ \text{s.t.}\ \nabla_{C_{q,p}} \sigma > 1 \}$: QPs falling into this interval have medium content relevance, indicating difficulty in ascertaining their content quality due to ambiguity. Different customers may exhibit very distinct engagement behaviors, resulting in high variance and reduced predictive power of engagement relevance. The transformation in this area, with the property $\nabla_{C_{q,p}} \sigma > 1$, magnifies the effect of content relevance, compensating for the weakened engagement predictability. This is another benefit of the sigmoid transformation.
    \item $R_3 = \{(q,p): c_2 \le C_{q,p} \le 1\ \text{s.t.}\ 0 < \nabla_{C_{q,p}} \sigma \le 1 \}$: QPs falling into this interval have high content relevance. They are flattened by the transformation and pushed to higher scores. The relative orders among these QPs are more determined by their engagement $E^{\mathcal G} \big(q, p\big)$.
\end{itemize}

The choices of $\alpha$ and $\beta$ determine ranges of $R_1$, $R_2$, and $R_3$. The two curves in Figure~\ref{fig:curve} exhibit the trade-off mainly between $R_1$ and $R_3$. The green curve, with a wide range of $R_3$, sets a low bar for a QP to be highly content-relevant, resulting in more QPs having high content scores. Conversely, the yellow curve, with a narrow range of $R_3$, sets a high bar for highly content-relevant QPs, leading to only a few of them being assigned high content scores.

In product search ranking, the top few positions, such as the top 5 and top 10, hold the most importance among the entire recall set, which usually contains hundreds or even thousands of products. This contrast emphasizes the need to prioritize those few items with the best content relevance in the top positions. Therefore, having a narrow but high-quality $R_3$ interval could potentially benefit the content relevance. This approach ensures the content quality of the $R_3$ products that are likely to occupy most of the top positions while still allowing products from other intervals to be promoted if they have considerably high engagement scores. As a result, rigorous criteria for products to be highly content-relevant could potentially improve the ranking model's performance in content-based relevance.

\subsection{Feature Generation}
\label{subsec:feature}

We employ a substantial number of features for ranking model training, with each feature being either engagement-related \cite{joachims2017user, zhao2021incorporating, sun2022personalized, qliu24engg, brand2022affinity, liu2022systems, store2024search} or content-related \cite{huang2013learning, magnani2022semantic, shan2023beyond}. For the content features, we use not only sparse features based on text match but also leverage LLMs to generate dense features. It is important to note a key difference in the size choice of language models for label vs. feature generation. Label generation can be performed completely offline where latency is not a concern, allowing us to use more complex LLMs to improve performance. However, features used in model training will also be computed at runtime during inference. While complex LLMs can deliver superior performance, their high computational cost presents a trade-off that must be carefully considered \cite{pradeep2021effectiveness}. Therefore, we must ensure that the size of the LLMs used to generate content features is kept within an appropriate range to avoid latency degradation.

Leveraging our rich in-house query and product attribute information and expert judgment, we train a moderate-size BERT-based \cite{bert} model with the cross-encoder \cite{xe} framework to generate content-related features for ranking model training and inference.

\section{Experiments and Evaluations}
\label{sec:experiments}

\begin{figure*}
  \includegraphics[width=\textwidth]{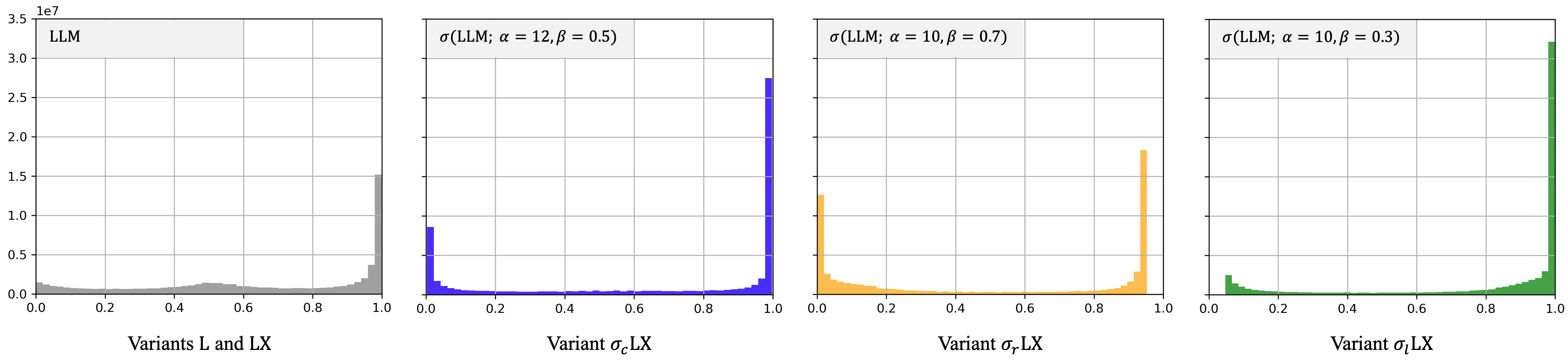}
  \caption{Distribution of LLM Scores and Transformed Scores in Histograms}
  \label{fig:dist}
\end{figure*}

With the proposed design in Section~\ref{sec:method}, we trained seven models, including one Baseline model and six Variant models as candidates. In the Baseline model, instead of using LLMs, we employ an XGBoost \cite{xgboost} model trained with a few content features to generate content relevance scores for labeling. Additionally, the Baseline model does not include the cross-encoder (XE) features in training or inference. In the Variant models, we apply LLMs to generate labels and/or include the cross-encoder features for training and inference. Model details are listed in Table~\ref{tab:models}. In Variants L and LX, we directly use the LLM-predicted value as the content-based relevance score as part of the label. In Variants $\sigma_c$LX, $\sigma_r$LX, and $\sigma_l$LX, we apply 3 different sigmoid transformations listed below.
\begin{itemize}
    \item \textbf{Variant $\sigma_c$LX} takes $\alpha=12$ and $\beta=0.5$, which only polarizes the high and low values without shifting the \textbf{\textit{center}}.
    \item \textbf{Variant $\sigma_r$LX} takes $\alpha=10$ and $\beta=0.7$, shifting the center to the \textbf{\textit{right}} while keeping $\sigma(0)$ and $\sigma(1)$ close to $0$ and $1$, without introducing excessively steep gradients in any sub-intervals. This sets a higher threshold, i.e., a more rigorous criterion for high content relevance.
    \item \textbf{Variant $\sigma_l$LX} takes $\alpha=10$ and $\beta=0.3$, shifting the center to the \textbf{\textit{left}} while keeping $\sigma(0)$ and $\sigma(1)$ close to $0$ and $1$, without introducing excessively steep gradients in any sub-intervals. This sets a lower threshold, i.e., a more relaxed criterion for high content relevance.
\end{itemize}
We plot the distribution of the original LLM-predicted scores and the transformed scores for all query-product pairs in the training data in Figure~\ref{fig:dist}.

\begin{table}
  \caption{Baseline and Variant Models in Experiments}
  \label{tab:models}
  \begin{tabular}{ccc}
  \toprule
  \multicolumn{1}{c}{\textbf{Model ID}} & \multicolumn{1}{c}{\textbf{Content Label}}                & \textbf{XE Features} \\ 
  \midrule
  Baseline                           & XGBoost                                                      & no                   \\
  Variant X                          & XGBoost                                                      & yes                  \\
  Variant L                          & LLM                                                          & no                   \\
  Variant LX                         & LLM                                                          & yes                  \\
  Variant $\sigma_c$LX               & $\sigma(\text{LLM}; \alpha=12, \beta=0.5)$                   & yes                  \\
  Variant $\sigma_r$LX               & $\sigma(\text{LLM}; \alpha=10, \beta=0.7)$                   & yes                  \\
  Variant $\sigma_l$LX               & $\sigma(\text{LLM}; \alpha=10, \beta=0.3)$                   & yes                  \\ 
  \bottomrule
  \end{tabular}
\end{table}

To measure the predictive performance of the Variant models, we evaluate both the content-based and engagement-based relevance of the ranking results generated by each model. The former is assessed through offline human judgment, while the latter is evaluated by conducting online interleaved A/B tests \cite{interleaving}.

\subsection{Offline Evaluation of Content Relevance}
\label{sec:eval}

To evaluate the content-based ranking relevance of the Variant models, we performed an offline human evaluation for each of them using NDCG metrics. The process starts with sampling a substantial number of representative search queries across all segments. For each query, we retrieve the top 10 products ranked by both the Baseline and Variant models. Human evaluators then assign 5-point relevance ratings to each query-product pair based on the product's relevance to the search query under well-defined guidelines. Finally, we calculate the NDCG@10 score for each query using the rating scores and item positions and conducted a T-test to compare the overall relevance quality between the Baseline and Variant models. 

The evaluation results are listed in Table~\ref{tab:offline_res}. The results lead to the following findings, which prove our hypotheses in Section~\ref{subsec:transformation}.

\begin{itemize}
    \item The comparison between Variants X, L, and LX suggests that simply adding XE features or incorporating LLMs in the label during model training alone does not result in a significant lift in ranking content relevance; however, combining these two approaches leads to a substantial improvement in content relevance.
    \item The comparison between Variants LX and $\sigma_c$LX indicates that the polarization used to push moderate relevances toward extremes during model training effectively enhances the model's predictive capability to prioritize highly content-relevant items in ranking.
    \item The comparison between Variants $\sigma_l$LX, $\sigma_c$LX, and $\sigma_r$LX suggests that, within an appropriate range, the more rigorous the criteria for high content relevance during model training, the greater the lift in content relevance achieved by the ranking model, and vice versa.
\end{itemize}

\begin{table}
  \caption{Content Relevance Evaluation Results}
  \label{tab:offline_res}
  \begin{tabular}{cc}
  \toprule
  \textbf{Variant Model}       & NDCG@10 Change (p-value) \\
  \midrule
  Variant X                    & $+0.41\%\ (0.11)$ \\
  Variant L                    & $+0.11\%\ (0.60)$ \\
  Variant LX                   & $\mathbf{+1.35}\%\ (0.00)$ \\
  Variant $\sigma_c$LX         & $\mathbf{+1.72}\%\ (0.01)$ \\
  Variant $\sigma_r$LX         & $\mathbf{+3.96}\%\ (0.00)$ \\
  Variant $\sigma_l$LX         & $-0.26\%\ (0.48)$ \\
  \bottomrule
  \end{tabular}
\end{table}

\subsection{Online Tests for Engagement Measure}
\label{sec:interleaving}

To measure the engagement performance of the Variant models, we conducted six interleaving tests, an online experiment where users are exposed to a mixed ranking list from both Baseline and Variant models on a substantial volume of online customer traffic at Walmart.com. The key metric is the percentage change in the number of items added to carts within the top 40 positions (ATC@40) for each Variant compared to the Baseline. 

The interleaving test results are listed in Table~\ref{tab:online_res}. The results suggest that using XE features and/or LLM labels does not significantly affect ranking engagement. A significant impact is observed only when applying shifted sigmoid transformations to the LLM labels, where engagement performance changes in the opposite direction to content-based relevance.

\begin{table}
  \caption{Engagement Relevance Test Results}
  \label{tab:online_res}
  \begin{tabular}{cc}
  \toprule
  \textbf{Variant Model}       & ATC@40 Change (p-value) \\
  \midrule
  Variant X                    & $-0.05\%\ (0.68)$ \\
  Variant L                    & $-0.06\%\ (0.60)$ \\
  Variant LX                   & $+0.05\%\ (0.11)$ \\
  Variant $\sigma_c$LX         & $+0.01\%\ (0.91)$ \\
  Variant $\sigma_r$LX         & $\mathbf{-0.79}\%\ (0.00)$ \\
  Variant $\sigma_l$LX         & $\mathbf{+1.04} \%\ (0.00)$ \\
  \bottomrule
  \end{tabular}
\end{table}

Specifically, comparing the performance of Variants ($\sigma_c$LX vs. $\sigma_r$LX), ($\sigma_c$LX vs. $\sigma_r$LX), and ($\sigma_r$LX vs. $\sigma_l$LX), we observe a pattern where an increase in content relevance is accompanied by a compromise in engagement. This suggests a trade-off between content-based and engagement-based ranking relevances, as anticipated in Section~\ref{subsec:label}. This observation highlights a pathway for search ranking model training: to construct a label incorporating both relevance aspects, apply appropriate transformations, and adjust parameter settings accordingly to achieve optimal performance toward specific relevance goals. As our selected candidates for balancing both relevances effectively, Variants LX and $\sigma_c$LX were further evaluated through comprehensive online A/B tests, and all business metrics, such as GMV and conversions, showed neutral results, validating that these models can achieve content relevance gains without compromising engagement.

\subsection{Root Cause Analysis}
\label{sec:rca}

To investigate the underlying causes of the observed evaluation/test results in depth, we establish a pathway by analyzing the relationship between feature importance, labels, and model performance. During model training, we calculate the averaged SHAP value \cite{lundberg2017unified} for each feature across all iterations and derive the ordered feature importance. Shown in Table~\ref{tab:featimp} is the feature importance rank and SHAP value of the highest cross-encoder LLM feature in different Variant models.

\begin{table}
  \caption{Feature Importance of Cross-Encoder LLM Feature}
  \label{tab:featimp}
  \begin{tabular}{ccc}
  \toprule
  \multicolumn{1}{c}{\textbf{Model ID}} & \multicolumn{1}{c}{\textbf{Importance Rank}}                & \textbf{SHAP value} \\ 
  \midrule
  Variant X                             & 8                                                           & 0.1141              \\
  Variant LX                            & 5                                                           & 0.2645              \\
  Variant $\sigma_c$LX                  & 3                                                           & 0.3817              \\
  Variant $\sigma_r$LX                  & 2                                                           & 0.6381              \\
  Variant $\sigma_l$LX                  & 7                                                           & 0.1808              \\ 
  \bottomrule
  \end{tabular}
\end{table}

We make four key noteworthy points here. Firstly, it is observed that utilizing LLM in labeling elevates the importance of the XE features in the model. This is because LLM enriches labels with more content-specific information, which better guides the content-related features to make more effective predictions. Consequently, the model relies more on these features during inference, resulting in more improved predictive capability to distinguish relevant from irrelevant products in ranking. Secondly, applying the sigmoid transformation further boosts the importance of the XE features due to its polarization effect. With more extreme content judgments embedded in label construction, the content-related features are able to make even more effective predictions, which further enhances the model's performance in determining content relevance. Thirdly, a more rigorous content criterion in labeling increases the importance of the XE features, while a more relaxed criterion reduces it. The reason is that greater rigor reduces the density of instances with high content relevance, making them more distinguishable, as shown in Figure~\ref{fig:dist}. This allows the content features to function more effectively in discrimination. Lastly, as the importance of XE features increases, engagement features become relatively less important, potentially causing a decline in customer engagement during inference. This explains the trade-off between content relevance and engagement relevance in performance from a feature perspective.

\section{Conclusion}
\label{sec:conclusion}

In this paper, we propose a novel approach to product search ranking model training centered on the integration of Large Language Models (LLMs) for both label formulation and feature generation. By leveraging fine-tuned LLMs alongside user interaction data, we optimize the model's training objective to account for both content-based and engagement-based relevance. The use of LLMs in labeling also enhances the effectiveness of LLM-driven content features in the ranking model, promoting more content-relevant products to top positions. Additionally, applying appropriate transformations to LLM scores in the labels further refines the balance between content and engagement relevances. Comprehensive online and offline evaluations demonstrate that our approach can yield significant improvements in content relevance while maintaining decent user engagement. This work provides valuable insights for enhancing training objectives, optimizing relevance, and implementing LLMs in e-commerce product search ranking.

%%
% The acknowledgments section.
\begin{acks}
We would like to express our gratitude to Nguyen Vo and Changsung Kang from the Walmart Search Ranking Team for their assistance with the language model training that supported this study.
\end{acks}

\bibliographystyle{ACM-Reference-Format}
\bibliography{MAIN}

\end{document}